\documentclass[sigconf, nonacm]{acmart}

\usepackage{xcolor}
\definecolor{DarkGreen}{HTML}{000000}
\newcommand\rr[1]{\textcolor{DarkGreen}{#1}}

\usepackage{graphicx}
\usepackage{caption}
\usepackage{subcaption}
\usepackage{todonotes}
\usepackage{multirow}
\usepackage{colortbl}
\definecolor{lightgray}{gray}{0.9}

\begin{document}

\title[Collaborative Co-Creation Process with AI]{Exploring the Collaborative Co-Creation Process with AI: A Case Study in Novice Music Production}
\author{Yue Fu}
\email{chrisfu@uw.edu}
\affiliation{%
  \institution{Information School, University of Washington}
  \city{Seattle}
  \state{Washington}
  \country{US}
}

\author{Michele Newman}
\email{mmn13@uw.edu}
\affiliation{%
  \institution{Information School, University of Washington}
  \city{Seattle}
  \state{Washington}
  \country{US}
}

\author{Lewis Going}
\email{lgoing7@uw.edu}
\affiliation{%
  \institution{Information School, University of Washington}
  \city{Seattle}
  \state{Washington}
  \country{US}
}

\author{Qiuzi Feng}
\email{qfeng5@uw.edu}
\affiliation{%
  \institution{Information School, University of Washington}
  \city{Seattle}
  \state{Washington}
  \country{US}
}

\author{Jin Ha Lee}
\email{jinhalee@uw.edu}
\affiliation{%
  \institution{Information School, University of Washington}
  \city{Seattle}
  \state{Washington}
  \country{US}
}

\renewcommand{\shortauthors}{Fu et al.}

\begin{abstract}
Artificial intelligence is reshaping creative domains, yet its co-creative processes, especially in group settings with novice users, remain under explored. To bridge this gap, we conducted a case study in a college-level course where nine undergraduate students were tasked with creating three original music tracks using AI tools over 10 weeks. The study spanned the entire creative journey from ideation to releasing these songs on Spotify. Participants leveraged AI for music and lyric production, cover art, and distribution. Our findings highlight how AI transforms creative workflows: accelerating ideation but compressing the traditional preparation stage, and requiring novices to navigate a challenging idea selection and validation phase. We also identified a new “collaging and refinement” stage, where participants creatively combined diverse AI-generated outputs into cohesive works. Furthermore, AI influenced group social dynamics and role division among human creators. Based on these insights, we propose the Human-AI Co-Creation Stage Model and the Human-AI Agency Model, offering new perspectives on collaborative co-creation with AI.

\end{abstract}

\begin{CCSXML}
<ccs2012>
   <concept>
       <concept_id>10003120.10003121.10011748</concept_id>
       <concept_desc>Human-centered computing~Empirical studies in HCI</concept_desc>
       <concept_significance>500</concept_significance>
       </concept>
 </ccs2012>
\end{CCSXML}

\ccsdesc[500]{Human-centered computing~Empirical studies in HCI}

\keywords{Human-AI Co-Creation, Creativity Model, Human-AI Interaction, Generative AI, Music, Novice, Agency}
\maketitle

\section{Introduction}
Artificial intelligence has afforded a new era of creativity, where humans and machines collaborate to produce novel and innovative works. Human-AI co-creation has emerged as a growing field of interest, capturing the attention of researchers and practitioners alike. The goal is to leverage the strengths of both humans and AI to produce creative outcomes that go beyond what either could achieve alone \cite{wingstrom2024redefining}. Such synergy has found applications across various domains, including writing \cite{Osone2021}, fashion design \cite{deng2024crossgai}, architecture \cite{tan2024using}, marketing and advertising \cite{islam2024transforming}, visual arts \cite{zhou2024generative}, and more.

\begin{figure*}[t]
    \centering
  \includegraphics[width=1.0\textwidth]{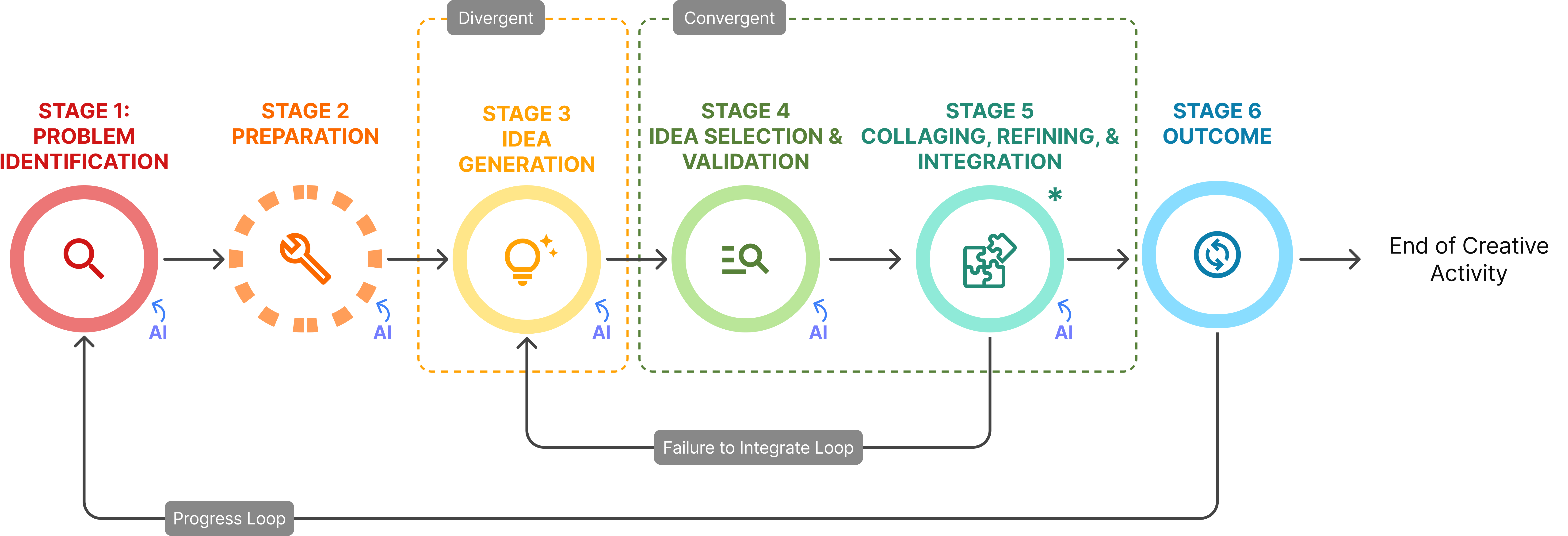}
  \caption{\textbf{The Human-AI Co-Creation Stage Model. }The diagram shows the creative process with AI support along a temporal dimension. As discussed in Section \ref{result_preparation}, introducing AI often shortens or bypasses Stage 2 (Preparation)---shown here with dashed boundaries. Although AI is praised for facilitating idea generation in Stage 3, it can also quickly lead to convergent stages in Stages 4 and 5, potentially linearize the creative process (see Section \ref{discussion_agency}). Additionally, we identify a new Stage 5 (Collaging, Refining, and Integration)---marked by an asterisk---where creators must merge AI-generated and human-created elements. Should this piece-together process fail, users revert to Stage 3 to regenerate compatible outputs. Involving AI also facilitate achieving tangible results, fosters a ``progress loop'' \cite{amabile2016dynamic}, giving creators a sense of accomplishment and prompting them to revisit Stage 1 for subsequent creation cycles. Finally, AI can be viewed as an environmental factor that supports user autonomy, competence, or task involvement throughout the stages, with the exception of the end outcome (see Section \ref{environmental_factor}). }
  \label{fig:stagemModel}
\end{figure*}

In music production, AI-powered tools have become increasingly popular, supporting both expert musicians \cite{BryanKinns2024, Banar2023} and novices \cite{Louie2020, KobayashiNoviceMusicDesign2024} in the creative process. A plethora of AI music generation and production tools powered by foundational models \cite{liu2023m} have been developed, aiming to facilitate music creation by increasing efficiency, enhancing self-expression, and supporting idea generation. AI music tools can compose melodies \cite{ju2021telemelody}, harmonize chords \cite{van2023algorithmic, tsushima2017function}, transcribe \cite{rigaud2016singing}, improvise \cite{madaghiele2021mingus}, suggest lyrics \cite{Wang2024}, and assist in mixing and mastering tracks \cite{Shtern2018}. For novice users, AI lowers barriers to entry, enabling individuals without formal training to engage in music production and express their creativity. Platforms like Suno \cite{sunoRef}, Google's Magenta \cite{magenta}, and OpenAI's MuseNet \cite{MuseNet} advertised they can democratize music creation by providing users with accessible tools that augment their creative capabilities.

Despite the growing prevalence of AI in creative fields, there is a notable gap in understanding the detailed processes of how creators collaboratively co-create with AI over extended periods. Much of the existing research has been conducted in laboratory settings or through online studies that capture only a snapshot of the creative process---often within a single session lasting less than an hour, using toy creative problems, and does not lead to final usable creative output. However, real-world creative artifact creation is typically a prolonged endeavor that involves various creative working stages and unfolds over weeks or months. In fields like architecture, design projects can span years, and in music production, though varying by genre and scope, it often takes weeks for a professional production team to produce a track from the initial concept to the final release.

Moreover, for much of creative processes, creation is a collaborative endeavor \cite{sawyer2017group}. In disciplines such as design, tools like Figma \cite{figma} and Miro \cite{miro} have revolutionized how teams work together, providing ways for people to collaboratively brainstorm, vote, and comment. Similarly, in music production, collaboration is the backbone of the creative process. It usually involves a team of individuals fulfilling various functions, including lyricists, composers, mixing and mastering engineers, and more. Each role contributes specialized expertise, and the interplay among team members is crucial to the success of the final outcome.
Using novice music production as a case study, our study aims to explore how AI tools influence collaborative dynamics and shape the stages of co-creation over an extended period. Music production represents an instructive example because it spans multiple creative roles (composition, arrangement, lyric writing, mixing/mastering) and often requires iterations to achieve a final outcome. Understanding both the collaborative dynamics and process of novice music production is essential for designing AI systems that effectively support collaborative co-creation activities between humans and AI. Specifically, we investigate three research questions: 

\begin{itemize}
    \item \textbf{RQ1:} What are the temporal stages of the co-creative process when novice users collaborate with AI in music production?
    \item \textbf{RQ2:} What relationships and roles do humans and AI assume in the collaborative music co-creation process?
    \item \textbf{RQ3: }How do users perceive AI in this process?
\end{itemize}

To address these questions, our case study explores how novice music creators work as a team to co-create music with AI over a period of 10 weeks. The setting of our study is an undergraduate capstone class at our institution, where students with no or limited previous music production experience used AI tools individually and collaboratively to create music pieces. They explored various AI technologies for different stages of music creation, including composition, arrangement, lyric writing, mixing and mastering, etc.

Our research team reviewed each team’s artifacts, including course submission materials, process logs, published songs, group websites, and final presentations. We then conducted in-depth interviews with nine participants to understand their experiences, the social and technical challenges they faced, and how AI shaped their creative and collaborative processes.

We found that AI significantly accelerated ideation but shortened the preparation stage---potentially limiting the deeper domain knowledge acquisition that often develops early in creative processes. The stage of idea selection and validation became both more critical and challenging, as participants were frequently overwhelmed by abundant AI-generated options with bounded evaluation knowledge. We also identified a new stage---collaging, refining, and integrating diverse AI outputs with human-created elements to form a cohesive song. Additionally, AI mediated social dynamics and boosted self-efficacy among team members, yet it sometimes constrained emotional expression and artistic exploration. Based on our results, we propose two models, the Human-AI Co-creation Stage Model and the Human-AI Agency Model, which offer new perspectives on collaborative co-creation with AI and provide design implications for enhancing both the creative process and agency.

Our study contributes to the understanding of collaborative human-AI co-creation in the following ways:

\begin{itemize}
    \item First, by focusing on a real-world, 10-week collaboration process, we capture the creative workflow across multiple stages and propose a \textbf{Human-AI Co-Creation Stage Model} that elucidates how creators and AI interact over time.
    \item Second, our analysis of collaborative teams, where members use AI both individually and collectively, reveals \textbf{AI tools mediate social dynamics and role assignment}.
    \item Third, by highlighting novice users' perceptions, we identify specific challenges and opportunities of AI-supported creativity in group settings. We then present a \textbf{Human-AI Agency Model} that argues for systems designed to nurture users’ emotional expression, skill growth, contextual flexibility, and community building.
\end{itemize}

\section{Related Work}
\subsection{Co-creation with AI in Music Production}
Co-creation with AI refers to collaborative creative processes where humans and AI systems work together, leveraging respective strengths to achieve a shared creative goal. Human-AI co-creation represents a fundamental shift in how we interact with technology, moving from a tool-based relationship to a collaborative partnership \cite{justThinkAI}.

Traditionally, music composition and production have relied heavily on human creativity and technical skill. Over the years, researchers in human-computer interaction and music education have explored this domain by developing digital musical tools---ranging from automatic mash-up systems and recording interfaces to gamified learning platforms and tutoring aids \cite{humphrey2013brief}.

With the advent of AI-powered systems, new possibilities have emerged to automate and augment various stages of music creation, sparking discussions about where these tools fit into the creative process \cite{Larsen2024, micchi2021keep} and how to ethically define roles and responsibilities for AI \cite{newman2023human, bown2021sociocultural, lee2022ethics}. In particular, AI has been proposed as a means to support novice creatives. For instance, neural networks based system complemented by steering tools can restrict generative notes to particular voices and nudge output in high level directions, thereby enhancing a novice’s sense of control and ownership \cite{Louie2020}. Additionally, novices often lack foundational music knowledge and are deterred by the complexity of traditional production tools, making them prone to reduced motivation \cite{Kobayashi2024}. Researchers emphasize tailoring AI solutions to novices' specific needs, preserving enjoyment and intrinsic motivation throughout the creative process \cite{Louie2020, Kobayashi2024}.

However, previous studies have predominantly examined individual musicians’ interactions with single-purpose AI tools in controlled settings \cite{louie2020novice, huang2019counterpointconvolution}, leaving a gap in understanding how teams of creators, particularly novices, integrate AI tools across an entire production cycle. As AI evolves from discrete utilities to collaborative partnership \cite{theberge2012end}, it is crucial to investigate how such technologies reshape collaborative processes. Our work explores how novice teams Co-create with AI tools over an extended music production period.

\subsection{Introduction of AI Tools Used in the Case Study }
Students in our study used a variety of AI tools. Here, we briefly introduce these tools and their functions to provide context for understanding our findings. 

Traditionally, there are various production elements in music production, such as composition (creating music content, such as melodies and chord progression), arrangement (organizing musical elements), recording (capturing audio), mixing (balancing and adjusting individual audio tracks), and mastering (optimizing overall sound and tonal balance)~\cite{senior2018mixing}. Digital Audio Workstations (DAWs) like Ableton Live \cite{ableton} serve as the modern central hub for music production, allowing creators to record, edit and manipulate both audio recordings and MIDI data (digital representations of musical notes)~\cite{bell2015can}. While DAWs have made professional-grade recording more accessible, they still require considerable technical skill to use effectively ~\cite{bell2015can,theberge2012end}. Each part requires significant technical expertise and musical knowledge, creating substantial barriers to entry for novice creators~\cite{senior2018mixing,bell2015can}.

Recent AI tools have emerged to support different aspects of the production process. For composition and arrangement, text-to-audio generators like MusicGen \cite{metaMusicGen} and Suno \cite{sunoRef} can create musical elements from text descriptions. Tools like Samplab \cite{samplab} can convert audio into editable MIDI data. For vocals, platforms like AIVA\cite{aiva} can create synthetic voices or train AI models on recorded human samples. AI mastering services like LANDR \cite{landr} use machine learning to automatically optimize and balance tracks. In our case study, students experimented with a variety of these AI tools, each targeting different parts of the production elements. Appendix \ref{Appendix:2} provides a detailed summary of the AI tools students experimented and used, alongside their core functionalities.

\subsection{Amabile’s Stage Model of Creativity---Our Theoretical Lens}
In the field of creativity research, scholars have proposed various models---such as the Four C Model \cite{kaufman2009beyond}, the Four P’s model \cite{gruszka20174p}, the Five A’s model \cite{gluaveanu2013rewriting}, as well as stage-based frameworks like design thinking and Pearlman’s model \cite{pearlman1983theoretical}---to capture the multifaceted nature of creative processes. Each framework offers unique insights into the origins, dynamics, and outcomes of creative work.

Amabile’s Componential and Stage Model \cite{amabile1983social, amabile1988model, amabile2011componential, amabile2016dynamic} is among the most widely recognized and cited in this domain. It explains how creativity emerges from the interplay of (1) domain-relevant skills, (2) creativity-relevant processes, and (3) task motivation. 
Beyond these components, Amabile’s model articulates five key stages of creative work: task identification, preparation, idea generation, idea validation, and outcome. These stages illustrate how individuals or small groups progress from recognizing a creative need to gathering information and skills, generating ideas, evaluating their viability, and deciding whether to accept a solution or loop back for further iteration. Additionally, Amabile emphasizes the role of the environment factors in fostering creativity, suggesting that supportive environment encourages risk-taking, autonomy, and open collaboration \cite{amabile1988model, amabile2016dynamic}.

We adopt Amabile’s stage model as our primary theoretical lens (see Section \ref{amabile_analysis}) due to its broad applicability across creative domains and its longstanding influence in creativity research. Our findings illustrate AI’s influences on each stage and need for a revision of the current model. Accordingly, we propose an updated version of Amabile’s framework that illustrates how involving AI fundamentally transforms people’s creative processes.

\section{Methods}
To explore how novice music creators co-create with AI tools over an extended period, we reviewed artifacts, process logs, and presentations by students who participated in a music production with AI course in the Spring of 2024. Additionally, we interviewed nine undergraduate students (out of 20 who attended the course). The interviews were conducted during the summer following the course completion, with each session lasting approximately 55 to 70 minutes. This study was reviewed by our institutional review board (IRB) and deemed exempt.

\begin{figure}[htbp]
    \centering
  \includegraphics[width=0.5\textwidth]{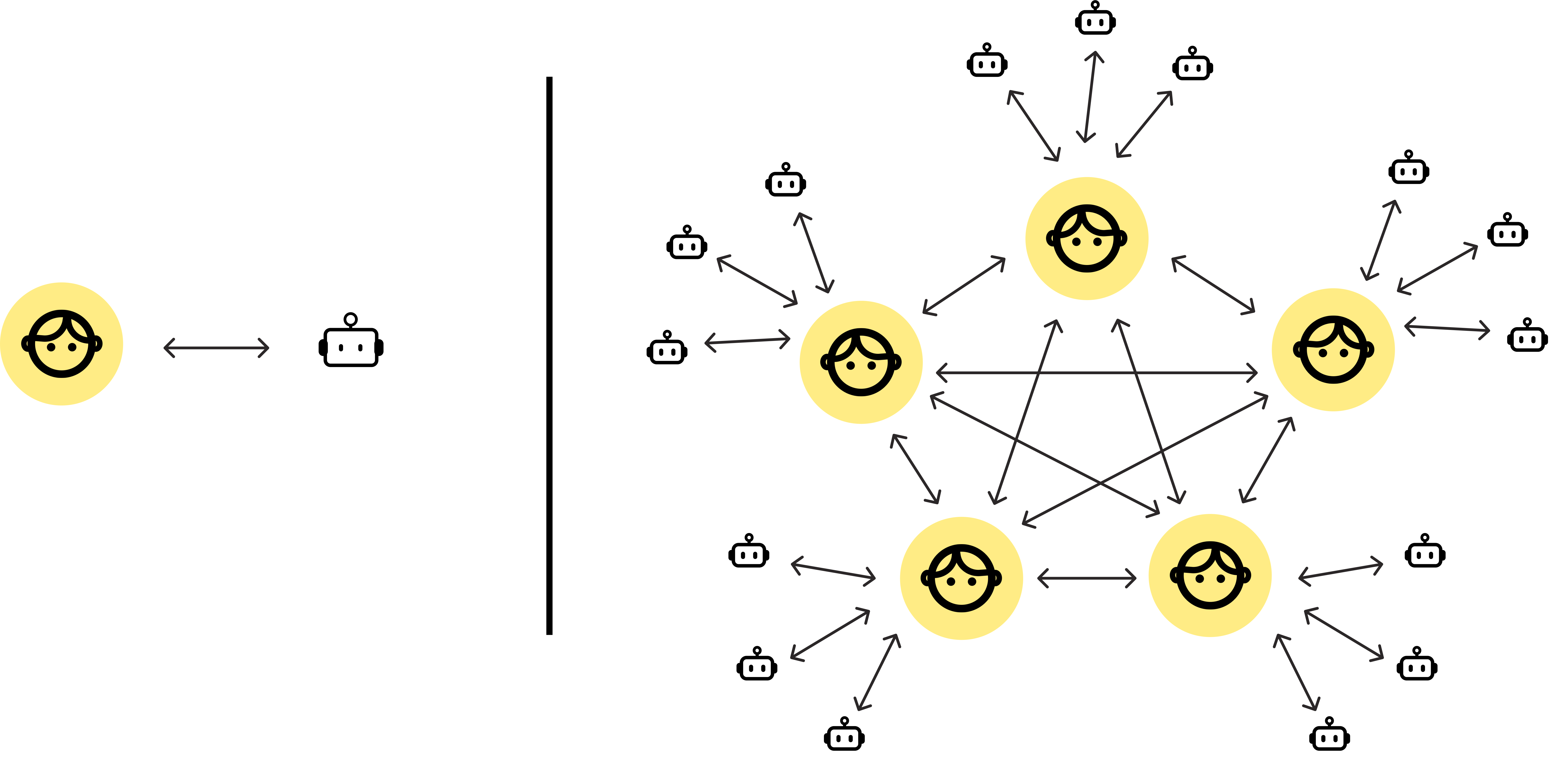}
  \caption{\textbf{Conceptual Illustration of Individual Versus Collaborative AI Co-Creation.} On the left, a single user interacts with an AI tool. On the right, it represents our case study context where students collaborated with each other and used various AI tools, reflecting a more complex and social situation.}
  \label{fig:collaborationModel}
\end{figure}
\subsection{Study Context}

The participants were enrolled in a 10-week capstone course at our academic institution, designed to let students explore AI and music creation in the digital age. The course aimed to let students experience through the entire music production process—from initial ideation to final distribution—culminating in the release of their music on platforms such as Spotify and students’ group websites.

Students were organized into teams of four to six members. Each team was tasked with collaboratively producing three tracks to be released as a single EP (see Figure \ref{fig:collaborationModel} for a conceptual illustration of students' collaborative co-creation context). The instructors provided an introduction of various AI tools in the beginning of the course. \rr{The introduction included a demonstration of MusicGen via Google Colab and provided students curated links of other AI tools. After this initial introduction, students were encouraged to explore independently and choose tools best aligned with their creative vision and project requirements.} Throughout the course, students used a variety of AI tools to support different stages of music production, including lyrics writing, composition, melody creation, instrument, mixing and mastering, and for distribution purposes, EP cover and website design. They also incorporated their own singing and vocal recordings, sometimes renting professional recording studios to mix their voices with AI-generated outputs. The course structure included lectures, weekly presentations of each team's progress, and a final presentation showcasing their EP and reflections on the process.

\subsection{Participants}

Participants were recruited through the course instructor and teaching assistant after the course had concluded. We reached out to all 20 students who took the course, and nine agreed to participate in the study. Prior to the interviews, participants completed an initial survey to collect demographic information and their prior music production experience. The nine students we interviewed belonged to three teams: four students from the first team (P1, P2, P3, P4 - Team A), three from the second team (P5, P6, P9 -Team B), and two from the third team (P7, P8 - Team C). See Appendix \ref{Appendix:1} for participant demographics. Each participant received a US \$50 Amazon gift card as a token of appreciation for their time. \rr{See Appendix \ref{Appendix:3}
 for the major AI tools each team used during the music production process.}
 
\subsection{Procedures and Materials}

We collected all the course homework materials from the instructor and teaching assistant, and reviewed the materials with participants' consent. \rr{This course provided a unique opportunity to examine the co-creation process systematically, as it explicitly required students to document their progress each week. These process materials encompassed iterations of group music concepts, pre-production plans, role assignments, ideation logs, reflections on experimenting with new AI tools, iterations of songs and lyrics, feedback from listening sessions, and participants' reflections on the feedback. The research team internally reviewed all these materials, including class presentations, collaborative online canvases (such as Figma), final lyrics, produced EPs available on Spotify, and each team's website for publicity and distribution. One research team member served as the teaching assistant for the course. This member provided a detailed, reflective document outlining students' experiences, challenges, processes, and specific AI tools explored during the class. This documentation was compiled from personal recollections and process logs submitted by students throughout the course.}

The research team met weekly to discuss and analyze the collected materials, identifying emergent themes and formulating potential research questions. Then three team members independently drafted semi-structured interview questions based on our agreed research questions. Over the next three weeks, the team discussed and refined interview questions, resulting in a final interview protocol comprising four main parts. The first part asked participants about collaboration dynamics. We explored the students' collaborative roles within their teams and their perceptions of AI's role in production. We also asked about the benefits and challenges faced when incorporating AI into a team setting. The second part asked about students' experiences with the music production process, focusing on how the involvement of AI technologies affected various music production stages and their individual workflow. The third part of the interview protocol explored participants' reflections on the 10-week music-making with AI journey. We asked their perceptions of how AI influenced their self-efficacy, creative expression, sense of ownership, and control and pride over the music pieces. \rr{During each interview session, the interviewer responsible for reviewing the team's material. Interviewers frequently incorporated excerpts from these materials, including listening recorded EPs together with interviewees, to prompt their specific memories and reflections.}

\rr{Most interviews were conducted in the summer to optimize participants' recollection of their experiences (the course was held during the spring quarter). Only one interview was conducted later after the summer break due to participant' availability.} Interviews were designed to last 60 minutes. All interviews were audio-recorded and transcribed. We anonymized and securely stored all audio recordings. 

\subsection{Data analysis}
\label{amabile_analysis}
We employed an abductive approach to analyze and understand the case by both existing related theories and empirical data collected. The abductive approach is often used in case studies  \cite{rashid2019case} and the process systematically combines theoretical framework, empirical data, and case analysis \cite{dubois2002systematic}. The aim of our abductive approach is to provide ideas and a subsequent tentative theory \cite{rashid2019case}.

We began analyzing the data and reviewing theories concurrently with the interviews. During the first round of coding, we used Reflexive Thematic Analysis \cite{dubois2002systematic}. One team member read the transcripts and revisited the audio recordings multiple times. The teammate extracted quotes and organized them into preliminary themes. The research team met weekly to discuss initial understandings, share insights, and compare interpretations of these quotes and themes. Through an iterative process, we agreed upon themes related to participants' usage, perceptions, and experiences of AI in the music production process. These themes encompassed various aspects such as collaboration dynamics, challenges and benefits of using AI, the influence of AI on creativity and workflow, and participants’ perceptions of AI in the music production process.

As our analysis progressed, we identified an alignment between the emerging themes and Amabile’s Model of Creativity \cite{amabile1988model}, which delineates creative stages. Recognizing this congruence, we integrated Amabile's model as a guiding framework to further structure and interpret our data. During subsequent meetings, the research team discussed how our data both supported and extended Amabile's model. We reached a consensus to adopt the theory for the next phase of analysis. This involved re-examining part of the data with Amabile's model in mind, and coding transcripts according to the creative stages of the model. We remained open to the possibility of modifying the model based on our findings.

The research team collaboratively refined the themes to reflect the expanded framework of creativity. Using an online collaborative tool (Figma), one researcher organized related transcript quotes into stages of creativity and human-AI collaboration themes. During a half-day data analysis workshop, four researchers collaboratively mapped participants' quotes, identified opportunities and challenges at each creative stage, and analyzed the workflow and iterations between stages in Figma. This workshop also focused on understanding human-AI collaboration within the creative process. Themes related to human-AI collaboration and participants’ perceptions of AI were documented and refined alongside the expanded creative stages. One researcher then synthesized the workshop findings in Figma, illustrating the expanded stages of creativity, directional links between stages, and overarching themes of human-AI collaboration. 
Subsequently, the lead author consolidated all the coded data, themes, and interpretations from the Figma board, as presented in the Results and Discussion sections.

\section{Results}

\subsection{Collaborative Co-creation Stages with AI}
\label{section:stage results}
\subsubsection{Stage 1: Problem and Task Presentation} \hfill

\vspace{0.65mm}
\noindent\textbf{Defining Creative Vision and Objectives.} In the initial stage, AI played little role. Participants focused on defining what they wanted to create, identifying the central theme, vision, and clarifying the message or “story” they aimed to convey. They discussed broad questions such as ``\textit{what kind of story [we] want to tell and the feeling [we] want the sound to resonate'' (P9).} One participant explained, ``\textit{we need to come up with the problem statement that we can explore different AI usage in the music production process}'' (P3), while another noted the importance of aligning their vision by understanding ``\textit{what we want to express [and] what we lack}'' (P4). These discussions not only guided their creative goal but also helped the teams form a collective motivation for the project.

Establishing shared goals and building consensus required active collaboration. Teams often held brainstorm sessions during which members presented ideas, preferences, and musical inspirations. These early meetings were critical for clarifying a \textit{big idea} or thesis for their music projects.

\vspace{0.65mm}
\noindent\textbf{Assigning Roles and Responsibilities.} Once the project vision began to crystallize, teams allocated tasks to specific members based on interests and skill sets. As one participant explained, ``\textit{We divide up the work at the beginning during the brainstorming stage…once we settled on a theme and genre, we [assign] the roles}'' (P7). Some teams assigned imaginary roles to AI, referring to AI as a ``\textit{teammate}'' (P4) and one group even gave it a playful name, ``\textit{Strawbae}'' (P8), which aligned with their band’s identity, signaling their intent to integrate AI as a meaningful player in this process.

We found this stage to be relatively brief, typically lasting only a single brainstorming or workshop session. During this time, students were assigned specific tasks that they would later support individually by using AI tools in the subsequent stages of the music production process.

\subsubsection{Stage 2: Preparation: Building Up Relevant Information and Skills} \hfill
\label{result_preparation}

In a traditional creative process, the preparation stage involves gathering knowledge, researching relevant information, and developing domain skills, which involves a great deal of learning and time-consuming \cite{amabile2018creativity, Bain1874-aa, poincare-cl}. However, in our case, students advanced quickly from Stage 1 (problem presentation) to Stage 3 (ideation). AI enabled fast idea generation without much preparation, reducing the need for knowledge-building at the outset.

When asked about their production process, none explicitly mentioned a distinct preparation or research phase. Their ``\textit{production plan}'' documents submitted as class assignments showed an immediate pivot from initial brainstorming to active ideation. While the 10-week capstone structure likely created external pressure to progress quickly, participants repeatedly noted that AI helped them dive into music creation with minimal prior expertise. Interestingly, several participants described using AI generation as a way to build domain knowledge. One student explained:
\begin{quote}
``\textit{We would create a sample using software [AI]. Then downloaded it, fed it into [an audio-to-text] AI tool. It [the tool] would read that music clip and generate text based on how it sounded... From there, we put that text into another tool to see what it would generate}'' (P8).
\end{quote}
Through feeding musical snippets into AI and receiving feedback or textual descriptions, students began to understand specific terminology and refine their vocabulary to prompt other AI systems. Another participant mentioned that an AI tool provides sample clips and patterns for the creator to generate and combine ideas, saying it ``\textit{helped [us] think creatively about what electronic or techno music we wanted to create}'' (P7), even though the final tracks did not incorporate that tool’s outputs.

Overall, we observed a blurring of the preparation and ideation stages. Rather than devoting significant time to mastering musical fundamentals, novice users relied on trial-and-error with AI to gain the knowledge needed for creation. This stands in contrast to the long “warm-up” periods often noted in creativity literature \cite{wallas1926art, simonton1999origins, amabile1983social}.

\subsubsection{Stage 3: Idea Generation} \hfill

The third stage of creativity involves generating ideas and exploring solutions. However, with AI, participants described a fundamentally altered process. Traditionally, individuals draw from their cognitive pathways and environmental resources \cite{amabile2018creativity}, but AI enabled them to produce music with less reliance on human cognitive and environmental resources. Our data suggest that AI served as a powerful springboard for inspiration, speeding up the ideation process, and a potential source of idea fixation on initial AI’s output.

\vspace{0.65mm}
\noindent\textbf{AI as a Springboard for Inspiration.} Many participants praised AI for jumpstarting their music making, even when they ultimately did not incorporate some of the AI-generated content in the final work. One said, ``\textit{you can get a lot of cool ideas, and especially for people who don't have any background in music. It does, I think, increase accessibility around how to create music or getting ideas}'' (P8). Another echoed, ``\textit{We ask ChatGPT to help us create lyrics on a certain theme…we might not use them directly, but they inspire us}'' (P7). Even when the team chose not to use the AI outputs, they perceived AI as useful. For example, one team generated musical clips via tools like Suno only to discard them while inspired by the ideas from the AI suggestions (P8).

\vspace{0.65mm}
\noindent\textbf{Direction Setting and the Risk of Idea Fixation.} Beyond providing inspiration, AI often guided participants’ musical direction. Several noted that ``\textit{it can quickly generate different types of songs with different styles…which sets the basis [for discussion]}'' (P4). Others found text-to-audio tools useful for ``\textit{getting a general vibe…[that] gave us some direction when we were lost}'' (P8). In many cases, the AI’s initial output shaped the foundational chords, beats, or melody line, and for some groups, effectively ``\textit{70 or 80 \% of the whole song}'' (P1).

However, this convenience came with a risk of over-reliance or idea fixation \cite{jansson1991design}. As one participant put it, AI may ``\textit{make you throw away ideas that could have been good if you had just developed them}'' (P6). Some team members acknowledged their limited knowledge, which prompted them to ``\textit{just go with what [AI] gave us}'' (P6). Novice status and AI’s “impressive” capabilities (P2, P4, P5) frequently led participants to adopt AI outputs without extensive exploration of alternatives.

\vspace{0.65mm}
\noindent\textbf{Accelerating the Ideation Workflow.} Participants repeatedly cited AI’s speed and efficiency in generating content. One student called AI ``\textit{really efficient…[it] can give us a whole song in maybe 30 seconds}'' (P3). Others described it as a ``\textit{shortcut}'' (P9) for tasks like chord generation and beat making. One student explained, ``\textit{It would have taken more time if we didn’t have AI…we were able to find chord progressions that fit well together}'' (P8). Specifically, participants repeatedly appraised AI’s ability to quickly generate lyrics (P2, P3, P5, P6, P8), beats and chord progressions (P1, P7), and simulate people’s singing voices (P1, P4, P7, P8).

\subsubsection{Stage 4: Idea Selection and Validation} \hfill

After generating a plethora of ideas generated by AI in Stage 3, participants moved on to selecting these ideas. As one student succinctly put it, \textit{``The output that AI generates needs to be selected by humans… The final choices are made by humans}'' (P1) Traditionally, individuals use domain-relevant skills acquired during Stage 2 to evaluate ideas for correctness or appropriateness \cite{amabile2018creativity}. However, because participants had little knowledge-building preparation, and AI could quickly produce numerous options, selecting and validating ideas became challenging.

\vspace{0.65mm}
\noindent \textbf{The Challenge of Validating Abundant AI-Generated Ideas.} Participants frequently felt overwhelmed by the sheer number of AI-generated options, finding it time-consuming to sift through them and decide which ideas merited further development. One participant recalled, ``\textit{We spent a large chunk of time trying to find something that we felt fit…And then it did become a little frustrating because we want to make progress on the song}'' (P8). Another participant mentioned generating a wide range of lyrics then need to filter out subpar content: ``\textit{That’s not to say I was blindly accepting AI stuff…a lot of the lyrics were just really corny}'' (P6). Lacking formal rubrics or deep domain expertise, many teams relied on gut feelings and intuition. One participant described hearing a promising AI-generated beat at the fifth iteration and disregarding the rest: \textit{``It just really resonate[d]… We couldn’t listen to the sixth or seventh iteration. We just stuck with that fifth one''} (P1). Another noted that, \textit{``}\textit{We tried a lot of stuff with AI but only a small fraction made it into the final song}'' (P9). For these novice producers, the subjective sense of what ``sounds good'' often served as the default standard.

While the abundance of outputs initially seemed advantageous, it often led to decision paralysis when students were faced with numerous choices \cite{schwartz2015paradox}, especially under the tight deadlines of a 10-week capstone course. One novice student noted their team was unable to appraise the various ideas generated by AI, ``\textit{At first, every idea was like a good idea. So we just tried to throw things together, but eventually, we realized that there were too many tracks and it was too cluttered.}'' (P6). Another participant noted that the selection phase was time-consuming, ``\textit{I think there was definitely a lot of inspiration we drew from what we generated using AI...But it was frustrating when we spent so much time.}'' (P8).

Time constraints yet prevented endless iteration. By limiting the available time for ideation, the course structure forced teams to converge on a final choice. As one participant mentioned, ``\textit{If there is no time limit, we will have endless discussions}'' (P4). Several participants recognized the need for a decisive leader to streamline the process, ``\textit{We have to arrive at a conclusion very quickly…[A] decisive person…makes the decision. We just used the idea with some adjustment}'' (P2). Thus external pressures and clear decision-making role can help manage the potentially overwhelming influx of AI-driven ideas.

\vspace{0.65mm}
\noindent \textbf{Balancing Individual vs. Collaborative Selection.} Teams differed in how they worked on idea selection and validation, with some emphasizing individual work before collectively deciding, while others favored a collaborative process from the outset. Certain participants iterated on AI outputs independently, ``\textit{I iterated many times…then chose a good one and showed it to the group}'' (P5), picking out viable lyrics or melodies from various generations on their own. In contrast, other teams tackled selection more communally. One group ``\textit{had a session where we each pulled up ten to twenty iterations and then shared them}'' (P7). Another group ``\textit{used Suno maybe 50 times and chose the best melody together}'' (P3). Whether individually or collectively driven, both approaches tried to reach a consensus efficiently and respected each teammate’s contribution.

\subsubsection{Stage 5: Collaging, Refining, and Integration with Human-Created Elements (The New Stage Caused by Involving AI)} \hfill

As a result of incorporating multiple AI tools and outputs into the creative process, participants introduced a new collaging and refinement stage in which they assembled materials from various AI sources into a cohesive musical work. In our case study, students typically found themselves working with scattered AI outputs often generated from different AI tools—melodies, chord progressions, vocal snippets, and sound effects—none of which fit together seamlessly. Additionally, they often needed to refine and adjust these fragments, then combine them with human-created elements such as raps, vocals, and instrumentation. In this stage, creators had to integrate an assortment of discrete pieces into a cohesive musical composition, much like stitching together diverse elements in a visual collage.

One student mentioned, ``\textit{everyone may use AI to do some different parts}'' (P4), and another explained, ``\textit{we used a lot of [AI tools], then we combined ideas… I actually spent [a great deal of] time pulling them together}'' (P2). The same student described in charge of putting teammates' AI-generated chords, beats, or lyrics and then using a digital audio workstation to merge and modify them: ``\textit{It starts with a melody…chords from another person…and lyrics from another teammate}'' (P2). P8 summarized succinctly, ``\textit{We blended the outputs}'' (P8).

\vspace{0.65mm}
\noindent\textbf{Necessity of Refinement and Adjustment.} Integrating AI outputs also demanded refinement of each AI output. One participant said, ``\textit{AI just produced the idea, and most importantly, it is our work in editing that}'' (P2). Another explained how they had to alter lyrics for better rhyme or tweak generated chords to suit the group's performance, ``\textit{If you want a comprehensive, emotional song, you have to do some refinements by yourself}'' (P3). Some misalignments, such as slight key differences or awkward transitions, forced participants to manually fix them to meet their musical standards (P2).

\vspace{0.65mm}
\noindent\textbf{Inflexible AI Audio Output.} Many teams found audio outputs from AI to be particularly challenging to work on. Participants often describe AI audio as ``\textit{fixed}'' (P5) and ``\textit{hard to manipulate}'' (P6). They pointed out that most AI systems only provide WAV or MP3 files, preventing users from easily adjusting individual notes or tracks. One student said, ``\textit{tools like that are really cool, but it's really hard to like deconstruct that and try to make something on your own with it because it's already like a finished product}'' (P6). Across \rr{participants}, students emphasized their desire for more control and flexibility over AI's output.

\vspace{0.65mm}
\noindent\textbf{Difficulty Mixing Human and AI Components.} In addition, when blending AI-generated segments with live instruments or vocals, participants ran into further complications: ``\textit{We tried to get a melody by integrating another instrument…that wouldn’t work because if you combine these two elements, it was just weird}'' (P3). Teams reported that once they had recorded human vocals, synchronizing or aligning with personal performance nuances was frustratingly complex. As one participant summed up, ``\textit{It’s really hard to incorporate [AI output] without it taking over the song…like 95\% AI and 5\% us}'' (P6).

\vspace{0.65mm}
\noindent\textbf{Time-Consuming Editing Processes.} Ultimately, assembling and refining these disparate pieces required significant time and skill. One participant described manually entering notes into Ableton, spending \textit{“four to five hours just editing the chords”} (P2). The same student observed that mismatches in AI's default musical keys forced students to retune or rewrite entire parts: ``\textit{It’s all black keys, but we sing in white keys, so it’s a serious problem}'' (P2). During this stage, considerable human effort is required to integrate into a polished final product.

Collectively, students often felt AI was falling short at this point in the process. While it could supply large quantities of raw material, integrating those diverse outputs into a polished track demanded significant human effort and expertise.

\subsubsection{Stage 6: Outcome} \hfill
\label{result_outcome}

Amabile's original model posits three possible outcomes for creative efforts: success, failure, or partial progress toward a goal \cite{amabile1983social}. In our study, AI mitigated the risk of ``no reasonable response possibility generated,'' \cite{amabile2018creativity} providing participants with “psychological safety” \cite{amabile2018creativity} for further exploration. As one student affirmed, ``\textit{I feel like we succeeded.} \textit{I don’t think I ever would have made a song if I didn’t have AI help me''} (P6). Even if the first song did not meet everyone’s exact expectations, AI frequently offered enough momentum to ensure progress. By the end of the course, each team produced and published three complete EP tracks---an accomplishment that students unanimously viewed as a success. The shared artifacts students created, including published EPs, helped them solicit feedback from peers, friends, and family. This feedback cycle reinforced what Amabile describes as a ``progress loop,'' \cite{amabile2016dynamic} whereby tangible advancement in meaningful work boosts intrinsic motivation and improves creative performance in subsequent iterations.

During this final stage, participants also developed more objective views of AI's limitations, particularly regarding the challenges of integrating and refining its outputs (see Stage 5). These insights prompted them to adapt how they would like to use AI in the next round. In addition, knowledge and experience gained from the initial trial is added to the participant’s domain skills. Over this process, students developed confidence and agency, as one remarked, ``\textit{Making an album with other people and AI sounded really hard at the beginning, but at the end, we found it wasn’t hard with teammates and AI}'' (P7). Notably, students often decide to rely more on their own rather than AI for future iterations. Collectively, our findings indicate that during this stage, students consciously reassessed the appropriate level of AI involvement, reevaluated their expressive needs, and integrated the domain expertise they had acquired before embarking on future iterations.

\subsection{Human-AI Collaboration}
Incorporating AI into the collaborative co-creation process reshaped social dynamics and influenced how students perceived their roles in the creative activity. While some teams treated AI as an additional ``teammate,'' participants acknowledged that AI differs fundamentally from human collaborators. This section details how AI mediated social interactions, how participants conceptualized AI's roles, why they valued human emotional input, and how they retained control over AI's contributions.

\vspace{0.65mm}
\subsubsection{AI-Mediated Social Dynamics} \hfill

\vspace{0.65mm}
\noindent\textbf{Sensitivity in Human-Human Collaboration.} Creating music as a team involves emotional risk and personal investment, making feedback exchanges especially delicate. One participant described learning ``\textit{when to step in and give feedback and when to let people do their own thing}'' (P6), highlighting the tension between honest critique and respecting others’ creative autonomy. Many admitted they were too polite to challenge ideas they did not like. ``\textit{People are a little hypocritical…nobody wanted to say bad things}'' (P5). Another participant observed how each member's different tastes demanded ongoing compromise: ``\textit{We each have different interests…there’s got to be some compromise there}'' (P6). Because most were still getting to know each other, they often defaulted to being ``\textit{receptive to everyone’s ideas}'' (P6), even at the cost of constructive critique.

\vspace{0.65mm}
\noindent\textbf{Emotional Ease and Honest Critique with AI.} In contrast, participants felt more comfortable offering blunt critiques of AI outputs. One student noted, ``\textit{It’s easy…you don’t have to worry about hurting AI’s feelings”} (P6). They routinely ignored or modified AI outputs generated by themselves or others without fearing interpersonal conflict. P6 explained, ``\textit{When your teammate shares something [they made]…you have to treat them gently, whereas with AI, it’s easy to ignore 90\% of what it says”} (P6). A different participant even credited AI for reducing social barriers around sensitive song topics, remarking that ``\textit{a judgment-free AI allowed us to bring walls down and execute on our vision}'' (P9). While AI was treated as a creative partner, its emotional neutrality made it easier to critique and reject its suggestions, a stark contrast to human-to-human interactions.

\subsubsection{Perceived Roles of AI in Music Production} \hfill

\vspace{0.65mm}
\noindent\textbf{AI as a ``Team Member'' and Co-Creator.} Some participants described AI as a creative partner that actively contributed to idea inspiration, lyric composition, and other roles. As one noted, ``\textit{It acted as both a songwriter and a lyric writer…like two group members at once}'' (P2). Another participant referred to AI as ``\textit{a fifth person in the room…someone to bounce ideas off of}'' (P6). Some teams even anthropomorphized the AI, giving it a name such as ``Strawbae'' to signal its perceived membership in the band. However, expectations of AI’s ability to maintain a team member role over time in creative collaboration often fell short. According to one student, ``\textit{We wanted Strawbae to be a band member, but…we found it was really difficult to give AI that kind of character}'', and by the end of the term, they had ``\textit{less confidence''} in it than they initially expected (P8).

\vspace{0.65mm}
\noindent\textbf{AI as Technical Assistant or Domain Expert.} Others viewed AI primarily as a powerful assistive tool, filling technical gaps where team members lacked expertise. One student believed AI should ``\textit{provide technical support and creative inspiration}'' (P3), while another admitted they ``\textit{regard AI as having more expertise than us}'' (P6), especially in mixing, mastering, or specific stylistic domains. Additionally, P2 admitted AI takes on tasks outside the participants' expertise, such as creating visuals for album art.

\subsubsection{Valuing Human’s Emotional and Creative Expression} \hfill

\vspace{0.65mm}
\noindent\textbf{AI’s Perceived Mechanical Output and Lack of Emotional Depth.} Participants frequently criticized AI-generated music as lacking the emotional depth that human-made compositions convey. One student commented, ``\textit{it lacks a certain amount of rawness you get from music made by humans}'' (P8). Another used a metaphor to describe AI outputs as ``\textit{too simple…like orange juice without added ingredients}'' (P2), highlighting their view that AI compositions do not capture the emotional nuance that resonates with listeners. A third participant recognized AI’s ``\textit{excellent technical processing,}'' yet lamented its inability to supply ``\textit{emotional expression, the space where human creativity remains irreplaceable}'' (P3). Others echoed similar sentiments, describing AI music as ``\textit{mechanical and stiff}'' or ``\textit{formulaic and monotonous}'' (P5).

\vspace{0.65mm}
\noindent\textbf{Importance of Emotional Connection.} Students underscored the emotional connection between listeners and creators. One participant explained that listeners ``\textit{want to be attached to the real people behind it}'' (P6), arguing that a composer’s lived experience or personal identity shapes how an audience emotionally engages with the work. Another student asked how one could meaningfully relate to an AI-generated piece: ``\textit{We relate symphony number five with Beethoven, his own experience and his own idea, his own expressions…But if this song is also written by a certain kind of AI model, maybe it reminds me of Beethoven, but it’s not really written by Beethoven. How can we relate this piece with the AI model itself?}'' (P1).

Several participants stressed the need to retain human emotional involvement, even as AI becomes more capable. One argued that ``\textit{final artistic expressions still need deep human involvement and emotional investment}'' (P3). One participant summarized succinctly, ``\textit{the optimal blend of AI and human creativity lies in finding a balance where AI's efficiency complimenting human emotional expressiveness}'' (P3), a stance reflecting the overall consensus.

\vspace{3cm}
\subsubsection{Value Ownership} \hfill

\rr{Participants expressed concerns around maintaining artistic ownership when collaborating with AI tools. The possibility of AI overpowering human contribution created a sense of unease.} One team explicitly preferred to use multiple, less powerful tools instead of relying heavily on a single, advanced ``\textit{Music AI" solution due to concerns that it might "verge on making its own song}'' (P6). \rr{Participants emphasized maintaining artistic ownership when collaborating with AI tools, as one participant articulated, ``\textit{I think the most important lesson I learned is to not let AI lead the way because music production in the end is like expressing your own identity and idea}" (P7). Another participant further echoed this perspective by asserting: ``\textit{I am expressing myself. It's my own manner. It might be different for other musicians, but my own manner is like, if AI is going to do it for me, it's AI's. It's not mine}'' (P1). This highlights a collective determination to preserve personal creative identities and avoid losing themselves to AI-driven processes.}

\subsubsection{A Conscious Choice to Retain Control} \hfill

\rr{Participants also demonstrated a strong and deliberate effort to retain control over the music production process. They consistently described actively engaging with, questioning, and editing AI-generated content rather than accepting it without critical review.} One student explained how they would ``\textit{question each line}'' in AI-created lyrics, ultimately scrapping many of them (P6). Another participant noted that ``\textit{AI generates better results, but at the end it's mainly us who wrote the prompts and chose…whether to use the end results}'' (P7), highlighting their conscious decision to maintain authority over the creative process. \rr{Over time, as participants became more experienced, they gravitated towards AI tools offering more adjustable parameters, thus providing greater flexibility and control.} They repeatedly emphasized that human goals and critical judgment should remain central, saying clearly that ``\textit{we can use AI to generate our idea, but never let it replace our thinking---we have to have our own thinking}" (P2). They concluded that humans must ultimately ``\textit{select}'' from AI outputs (P1), ensuring their control of the process and outcome.

\subsection{AI's Impact on Creative Expression and Self-efficacy}

Participants expressed both positive and negative views regarding AI's influence on their creative expression, praising its ability to boost confidence while also critiquing its constraints on artistry and collaboration.

\subsubsection{Enhancing Creative Confidence and Self-Efficacy} \hfill

Many students credited AI with bolstering their creative confidence and enabling them to explore musical ideas they might not have attempted otherwise. One participant acknowledged they ``\textit{never would have made a song}'' without the reassurance provided by AI (P6), explaining that simply knowing there was a tool to fill gaps in their knowledge empowered them to experiment their ideas. Another described the experience of using AI as ``\textit{pushing me outside my comfort zone,}'' which eventually led to personal growth and a willingness to try to produce music in the future, ``\textit{if that's not like what I'm gonna do for the rest of my life, it could still be like a fun side hobby}'' (P8). For newcomers to music production, AI mitigated the ``\textit{cold start}'' problem and made the creative process more inviting (P7). Consequently, many felt a boost in self-efficacy, taking pride in what they had achieved and viewing the experience as a stepping stone for future creative endeavors.

\subsubsection{Potential Constraints on Artistic Expression} \hfill

However, participants with stronger musical backgrounds or a desire for deeper artistic control sometimes viewed AI as limiting their creative expression. One student lamented that relying on Suno to generate entire songs reduced opportunities for collaboration and hands-on experimentation, leading them to ``\textit{spend less time with each other…or with real physical instruments}'' (P7). Others found that constantly regenerating AI outputs became ``\textit{overwhelming and disruptive}'' (P1), sapping their motivation to pursue personal melodies or ideas. In some cases, frustration with subpar or repetitive AI content generation caused teams to ``\textit{spend so much time}'' trying to salvage material that ultimately failed to meet their artistic vision (P8). While AI undoubtedly boosted confidence for novices, it could also stifle the creative instincts of those who wished for a more human-driven and embodied approach to music-making.

\section{Discussion}
\subsection{Design Implications for the Human-AI Co-Creation Stage Model}
In this subsection, we connect the results in Section \ref{section:stage results} and propose a Human-AI Co-creation Stage Model (see Figure\ref{fig:stagemModel}). The model builds on and extends Amabile's original model of creativity \cite{amabile1983social}. We also discuss the challenges and opportunities AI offer and propose a list of design implications for each stage of the co-creation process.

\subsubsection{Protean Roles of AI and Supporting Initial Problem Space Exploration} \hfill
\label{environmental_factor}

Many students in our study were initially uncertain about AI’s potential, often personifying it as a ``fifth teammate''. However, AI is neither human nor strictly confined to any single role \cite{vartiainen2025emerging}. Instead, it can perform multiple functions---from generating chord progressions to synthesizing vocals in our case. Viewing AI as an environmental factor \cite{amabile2016dynamic, newman2024want} or resource, rather than a discrete collaborator, may offer a more adaptable and efficient framework for creativity. \textbf{Environmental factors} refer to external influences that either support or hinder the creative process. Using this perspective, the presence of AI can impact an individual's intrinsic motivation, access to resources, and ability to engage in creative thinking. 

\rr{AI systems, particularly those designed for novice creators, could significantly enhance initial exploration of problem and solution spaces. Novices, limited by their domain knowledge, often struggle with effectively navigating creative problem spaces and identifying directions. AI system designers should consider mechanisms that \textbf{externalize and illuminate the creative problem spaces }for users, providing structured guidance or exploration cues. For example, systems could incorporate high-level overviews of possible musical genres, thematic prompts, or even visual inspiration in the form of preliminary mood boards, thereby helping novice creators systematically scope and understand the creative possibilities at their disposal.}

\rr{Moreover, incorporating \textbf{multi-agent AI systems that simulate various persona roles} can further enrich the initial problem-space exploration by introducing multiple perspectives. Such multi-agent systems, as demonstrated in recent collaborative AI frameworks \cite{yes_and}, can foster diverse dialogues that human collaborators may not naturally consider due to limitations in their experience or domain-specific knowledge. Encouraging users or creative teams to interact with multiple AI personas rather than assigning a singular, narrowly defined function to AI may thus facilitate richer, more multidimensional problem space exploration. }

\subsubsection{Supporting Preparation and Domain Knowledge Building} \hfill

In our study, the Preparation stage (Stage 2) was often compressed or entirely bypassed, with participants swiftly transitioning from Task Identification (Stage 1) to Idea Generation (Stage 3). We attribute this rapid progression to the availability of powerful AI tools that facilitate quick ideation, thereby reducing the perceived need for an extensive ``warm-up'' or preparatory period. However, the preparation phase and the development of domain knowledge are critical not only for understanding the creative activity initially but also for effectively evaluating ideas during Idea Selection and Validation (Stage 4). The absence of a thorough preparation stage may undermine creative outcomes, as creators under external pressure might favor more predictable, AI-generated solutions without sufficient domain expertise \cite{amabile2018creativity}.

Despite these challenges, there is significant potential for AI to support more systematic knowledge and skill exploration. Foundational AI models can provide domain-specific knowledge, sense-making support, and contextual explanations tailored to novices’ needs. \rr{Designers should therefore \textbf{integrate learning objectives} within AI systems, especially when these systems are expected to serve large novice user populations. } For example, in the domain of image generation, tools like CreativeConnect \cite{CreativeConnect} enable users to analyze reference images, break down suggested keywords into categories, and recombine elements to create new images with AI assistance. By delivering contextual explanations and tailored domain knowledge, AI can provide a structured preparation phase before engaging in ideation. 

Moreover, \textbf{intentionally delaying the introduction} of the AI-driven ideation phase to incorporate a ``training the brain'' period may foster deeper cognitive engagement and mitigate the risk of early cognitive fixation on suboptimal ideas \cite{guo2023rethinking}.
\rr{Aligning this design approach with educational theories of ``productive struggle" \cite{warshauer2015productive}, it is beneficial for users to grapple with challenging problems during this stage before accessing AI-generated ideas. By deliberately delaying immediate AI idea generation, designers can encourage deeper cognitive engagement, mitigate the risks associated with overreliance, and reduce anchoring biases.}
\subsubsection{Supporting Divergent Inspiration and Preventing Convergent Fixation } \hfill

Students in our study consistently praised AI for its ability to rapidly generate numerous creative ideas, serving as a powerful tool for inspiration and exploration. As multiple studies have noted, AI systems can function effectively as brainstorming aids \cite{yu2023investigating, shaer2024ai, kim2023effect}, offering a seemingly endless array of possibilities. However, this capability also tends to drive an early convergence in the ideation process. Traditionally, creative ideas emerge gradually, often following a prolonged incubation phase during which vague ideas evolve into clear, innovative concepts \cite{wallas1926art}. In contrast, AI frequently delivers near-final or highly refined outputs. This immediate availability of polished ideas can narrow a creator’s exploratory range, leading them to focus on \textit{perfecting specific prompts} rather than venturing into broader, alternative creative pathways. \rr{To avoid this pitfall, it becomes crucial for AI system designers to \textbf{actively encourage idea divergence} rather than merely providing the most straightforward or obvious ideas. Potential design features such as `\textbf{`big-picture'' overviews} of stylistic or thematic \textbf{variations} can encourage ongoing divergence and prevent users from locking onto the first promising AI-generated concept. Also, AI systems that prompt reflection on initial goals (Stage 1)  can support users' metacognition and help them stay open to broader creative directions.}

\rr{Additionally, inspired by human practices in art and design, introducing \textbf{intentional ambiguity} into AI outputs can serve as a resource for imagination and interpretation \cite{gaver2003ambiguity}. Rather than presenting polished and fully realized ideas, AI systems could offer incomplete or ambiguous ones, such as rough narratives or metaphorical prompts, to intentionally invite human creators to fill in the gaps, select resonant ideas, and infuse personal meaning. In practice, AI systems might feature multiple rough drafts or even a ``wild ideas" mode, explicitly signaling to users that these suggestions are intended as creative seeds.}

AI-enabled creativity also shifts the burden from generating ideas, which traditionally a significant obstacle, to selecting and evaluating them. Participants in our study expressed decision fatigue \cite{hirshleifer2019decision} when confronted with numerous AI-generated options. Without extensive domain knowledge or preparation, they often defaulted to intuitive judgments, such as ``it sounds good''. Creators may gravitate toward a \textit{satisficing} choice, selecting an adequate option rather than striving for the optimal one, an example of bounded rationality \cite{simon1957behavioral}. Being highly subjective, music domain may validate such a gut-level approach. However, over-reliance on intuition risks cognitive biases, akin to System 1 thinking overshadowing more deliberate System 2 reasoning \cite{kahneman2011thinking} and intuition judgment may not apply to other co-creation domains such as marketing and architecture design. To address this, AI tools could integrate \textbf{evaluation guidelines and metrics}, or offer scaffolding that helps creators formulate their own. By clarifying the criteria for ``success,'' such features help users more accurately gauge the quality and relevance of AI outputs. Also, \textbf{automated curation features might filter out obviously low-quality outputs} and provide explanations for each dismissal, reducing cognitive overload. Clustering or categorizing similar ideas further streamlines comparison and selection, allowing users to focus on the most promising concepts and expand their domain knowledge.

\subsubsection{Challenges in Collaging, Refining, and Integration} \hfill

Our findings reveal a new stage that extends beyond Amabile’s original framework: Stage 5, focused on collaging, refining, and integrating AI outputs with human-produced material. Many described the process as time-consuming and technically challenging, with AI outputs often arriving in uneditable formats that impede integration. Consequently, promising ideas are sometimes abandoned simply because they cannot be made to ``fit'' with other elements.

To address these integration challenges, AI systems should aim to produce more \textbf{flexible and easily editable outputs}. Specifically, designers could design AI-generated content to be available in formats that facilitate detailed adjustments---such as separate MIDI tracks, discrete vocal stems, or clearly defined audio layers. Incorporating \textbf{real-time or in-situ editing capabilities} directly within the AI generation interface could enhance workflow efficiency and creative control. Moreover, it is critical to streamline and optimize the workflow pipeline between different AI tools and digital audio workstations. Seamless export-import functionalities should be prioritized, allowing AI outputs to transition smoothly into DAWs or other software environments to reduce the manual overhead..

\subsubsection{Sustaining Progress and Motivation} \hfill

Our results (Section \ref{result_outcome}) show all groups completed an EP consisting of three songs, demonstrating involving AI leads to success or at least help human creators gain some progress towards creative goal established in Stage 1 (Task Presentation). By achieving tangible results, creators experience a ``progress loop'' \cite{amabile2016dynamic}, where a sense of accomplishment boost motivation and confidence, prompting them to loop back to Stage 1 for another cycle of ideation. This positive feedback cycle can foster high levels of creativity over multiple iterations.

AI system designers should, therefore, consider mechanisms that extend beyond facilitating content production to \textbf{actively supporting sharing, feedback, and broader social interactions}. \rr{As indicated by recent developments in the generative AI field---such as Suno CEO's vision of seeing Suno as a form of social media \cite{suno_interview}, and OpenAI's plans to develop social media capabilities potentially influenced by GPT-4o’s successful image generation feature \cite{4oImage}---the social dimensions of creative AI platforms are increasingly recognized as central to user motivation and engagement. Sharing creative products with others can be intrinsically rewarding, heightening creators' sense of accomplishment and promoting further exploration and deeper involvement with creative tasks. }

Moreover, providing \textbf{reflective evaluation} features at the final outcome stage could further sustain creators' motivation by helping them critically assess both human and AI contributions to their completed works. Such reflective tools empower creators to make informed decisions about their subsequent use of AI, ensuring future engagements align closely with their evolving creative goals and personal preferences. 

\subsection{AI-mediated Social Dynamics and Roles}
Creativity is often framed as a deeply social process \cite{kaufman2010cambridge}, especially in domains like music where improvisation, composition, and performance traditionally hinge on collaboration and interpersonal exchange. The creative collaboration has been proposed to be  ``distributed'' among audiences, materials, embodied actions, and the socio-cultural environment, which expand the collaboration concept to beyond direct human collaboration \cite{barrett2021creative}. Our data shows AI can assume dual functions within a creative team---it can appear as a ``teammate'' to whom creators assign a name (e.g., ``\textit{Strawbae}''), or it can serve as a environmental factor that shapes the creative environment without behaving as a conventional collaborator, serving as technical tools and socio-cultural contexts.

These AI-mediated social dynamics influenced how participants critiqued work and navigated interpersonal relationships. While students often hesitated to offer candid, potentially harsh feedback on their peers' contributions, they felt no such inhibitions about criticizing AI output---even when it was co-created by a human teammate. This tendency resonates with findings from AI-mediated communication research \cite{hancock2020ai, fu2024text}, where research shows individuals partly blame AI rather than attributing all responsibility to their conversation partners \cite{hohenstein2020ai}. In group settings, AI can thus function as a buffer or scapegoat, easing interpersonal tensions by absorbing some share of blame or critique. \rr{To further facilitate productive critique while managing interpersonal tensions, platform designers could consider implementing collaborative systems that \textbf{enable anonymous and equitable feedback}. Such anonymity can mitigate social pressures and biases, fostering honest evaluation of both human and AI-generated content, and contributing to healthier collaborative dynamics. Another approach to facilitate collaboration involves explicitly positioning AI as a \textbf{mediating agent }within social interactions. Such mediating roles for AI may enhance team cohesion, providing a balanced distribution of social accountability. }

Our study also highlights how AI-driven successes may \textbf{foster inflated self-assessment} among novice creators. Many participants lauded their final works as their own achievements, even if AI played a decisive role in shaping the music or lyrics. This self-attributed success can cultivate optimism and self-efficacy, motivating further experimentation with AI. However, this self-attribution may be somewhat illusory, given that the compressed preparation phase in our study might have left creators with limited domain knowledge, thereby constraining their ability to critically appraise the quality of their outputs.

\begin{figure}[htbp]
    \centering
  \includegraphics[width=0.45\textwidth]{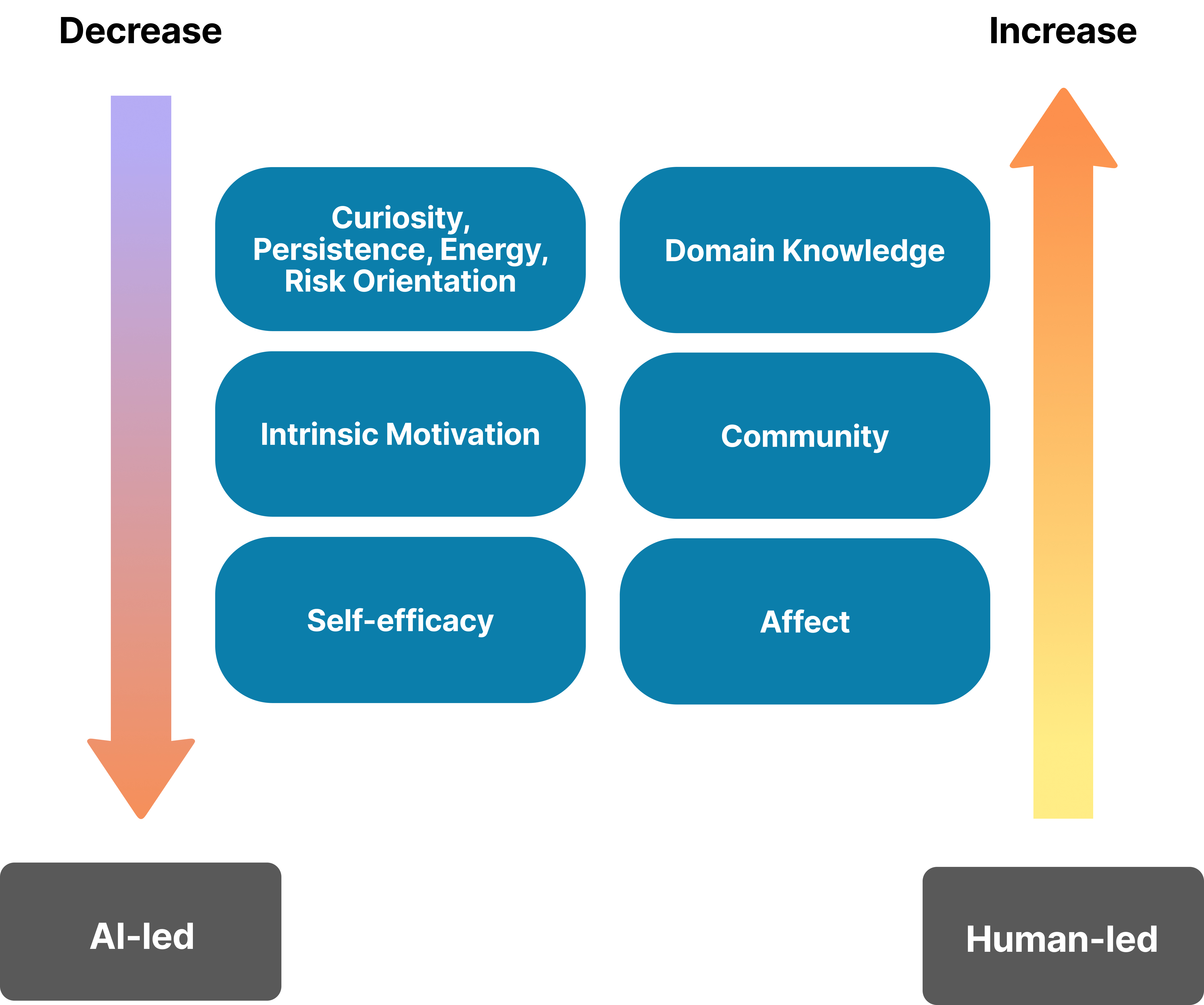}
  \caption{\textbf{The Human-AI Agency Model.} AI-led creation tends to lower motivation, self-efficacy, and domain growth, etc., whereas human-led processes can increase these factors, supporting creators' long-term growth and intrinsic enjoyment of creativity (see Section \ref{discussion_agency}) for discussion.}
  \label{fig:agencyModel}
\end{figure}

\subsection{Human-AI Agency Model in Co-Creation}
\label{discussion_agency}
We propose a Human-AI Agency Model (see Figure \ref{fig:agencyModel}) that positions users in the driver’s seat, promoting not only better creative outcomes but also a deeper sense of intrinsic value and enjoyment throughout the creative process. Historically, creative activities (e.g., art-making, music-making, research) have intrinsic values that are revealed from repeated engagement, not solely from the quality of the final artifact \cite{zhang2024searching}. While AI can greatly augment the creative process, particularly for novices lacking the skills to begin, it can also detract from the enjoyment and sense of agency inherent in creativity.

One-click AI generation tools risk diminishing human agency, stripping away the exploratory ``aha'' moments. Moreover, while the positive affect associated with early AI successes encouraged broader cognitive exploration, one-click generation tools risk reducing creators to mere prompt engineers---limiting deeper engagement, collaboration, emergence, and the nuanced evolution of their creative visions. Consequently, using AI to support creative activity may also lead to a more \textit{linearized} workflow: following specific process of prompting, validating, and refining. This tension underscores the importance of contextual flexibility---helping creators harness AI’s efficiency while preserving the open-ended, playful spirit that fosters exploration and serendipitous discoveries.

Additionally, participants consistently emphasized that certain aspects of music-making, particularly emotional expression, should not be replaced by AI. This resonates with broader concerns about AI-generated media and the risk of fabricating emotion or persona \cite{bakir2024manipulation, yonck2020heart}. While an audience might not differentiate between human and AI-driven musical output, the intrinsic reward of conveying personal feelings and building a human-to-human connection is central to the creative experience. AI system designers should focusing on supporting people's emotional expression but not using AI to automate and replace human input.

We also argue that it is important to think beyond the transactional use and immediate outputs \cite{newmanpurposeful} of the AI tools and think about how these tools can support humans to grow as a creator over time. This includes nurturing curiosity, honing heuristics, and building up intrinsic motivation. An ideal system might adopt an ``easy to use, hard to master'' model, offering novices a gentle learning curve while allowing experienced users to refine more sophisticated skills. By reinforcing self-efficacy and providing structured pathways for skill acquisition, AI can act not just as an accelerant for creativity but as a coach-like guide.

Finally, much of today’s AI co-creation revolves around individual interactions. Yet music production, and many other creative fields, often thrive on collaboration and social sharing. Future AI designs could integrate multi-user environments, enabling teams to compose and refine content together, share updates, and build social communities around their work. These environments can facilitate \textit{publication and feedback loops}, allowing creators to share outputs with peers, family, or broader audiences for continuous dialogue and engagement.

\rr{Our findings suggest several key implications for the design of human-AI co-creation systems. First, designers should \textbf{facilitate controlled serendipity} \cite{andersen2016conversations} and \textbf{contextual flexibility}, encouraging non-linear workflows that allow broad experimentation and support spontaneous, cognitively emergent discoveries. Second, platforms should explicitly \textbf{foster growth-oriented skill development} by integrating tools to track user progress, providing tailored tutorials, contextual tips, and adjustable challenges to sustain long-term mastery and intrinsic motivation. An ideal co-creative AI system might employ an ``\textbf{easy to use, hard to master}" approach, inspired by game design principles, offering accessible interfaces for novices while gradually revealing advanced, nuanced controls as users develop proficiency. Third, AI systems should actively \textbf{support emotional expression without replacing it}, avoiding simplistic automation of expressive tasks, and instead enabling creators to incorporate their personal stories, imagery, or poetic expressions beyond basic emotional cues. Finally, considering the inherently social dimension of creativity, AI design should \textbf{facilitate collaborative co-creation and social sharing}, expanding beyond single-user scenarios to multi-user environments. Such platforms would help enable real-time collaboration, publishing, and iterative feedback, fostering vibrant creative communities and sustained user engagement.}

\section{Limitations and Future Work}

Our study has several limitations. First, we have a limited number of participants involved in the interview process. Second, the structured course---with its assignment deadlines and the requirement to produce three songs---introduced external pressures that may have constrained participants’ creative processes relative to real-world music production. Third, the rapid formation of student teams in the beginning of the course likely inhibited the development of the deep interpersonal rapport typical in long-term professional collaborations. Future work should extend this research to larger and more diverse populations, examine longitudinal creative collaborations in professional or community settings, and explore how advanced, domain-specific AI systems integrate into extended, real-world co-creation processes.

\section{Conclusion}
Our study explored how novice music creators collaboratively co-create music with AI over a 10‐week period. We found that while AI accelerates idea generation, lowers barriers to entry, and boosts novices’ confidence, it also introduces unique challenges. Notably, AI involvement compresses the traditional preparation stage and gives rise to a new collaging and refinement stage. Furthermore, AI-mediated social dynamics alter how students critique one another’s work and negotiate roles. By situating our findings within an updated Human-AI Co-Creation Stage Model, we demonstrate how AI influences every stage of the creative process---from initial task identification to final outcome assessment. In addition, we propose a Human-AI Agency Model that emphasizes the need to preserve human emotional input and argue for intrinsic value of creative activities. We suggest future human-AI co-creation systems should foster creativity, collaboration, and skill development,  enhancing rather than eclipsing the essential human experiences of artistic expression.

\bibliographystyle{ACM-Reference-Format}
\bibliography{references}
\appendix
\section{Participant Table}
\label{Appendix:1}
\begin{table*}[ht]
\centering
\caption{Participant Demographics, including team assignment, prior music production experience, prior use of AI tools supporting music production, and reported gender.}
\label{table:participant_demographics}
\begin{tabular}{lllll}
\toprule
\textbf{PID} & \textbf{Team} & \textbf{Music Production Experience}& \textbf{Used AI Production Tools}& \textbf{Gender} \\
\midrule
P1 & Team A & < 1 year & No & Male \\
P2 & Team A & 1-2 year & No & Male \\
P3 & Team A & No experience & No & Female \\
P4 & Team A & No experience & No & Male \\
P5 & Team B & No experience & No & Male \\
P6 & Team B & No experience & No & Female \\
P7 & Team C & 1-2 year & No & Female \\
P8 & Team C & No experience & No & Female \\
P9 & Team B & < 1 year & No & Male \\
\bottomrule
\end{tabular}
\end{table*}

\section{Tools For Music Production Tried and Used During the Course}
\label{Appendix:2}
Students tried and used a range of AI tools. These tools can be categorized based on their primary function:

\textbf{Text-to-Audio Generation:}
\begin{itemize}
    \item Tools like Meta's MusicGen, Google's AudioLM, and OpenAI's Jukebox.
    \item These tools use natural language processing and deep learning to interpret text descriptions and generate corresponding audio.
    \item Output quality and length can vary. For example, MusicGen typically produces 15-second clips.
    \item Student use cases include creating custom sound effects, background music, or starting points for compositions.

\end{itemize}

\textbf{Audio-and-Text-to-Audio Generation:}
\begin{itemize}
    \item MusicGen and similar models like AIVA or Amper Music offer this functionality.
    \item This approach combines the benefits of text prompts with audio references, allowing for more precise control over the output.
    \item The audio input can serve as a style guide, tempo reference, or melodic inspiration.

\end{itemize}

\textbf{Audio-to-Audio Generation:}
\begin{itemize}
    \item Tools like Magenta Studio Extend, AIVA, and Splash Pro use this technique.
    \item These models can extend existing audio clips, generate variations, or create complementary tracks.
    \item Useful for expanding short loops, creating B-sections for songs, or generating alternative arrangements.
\end{itemize}

\textbf{Audio-To-MIDI Generation:}
\begin{itemize}
    \item Platforms like Samplab, Melodyne, and AudioToMIDI use this technology.
    \item These tools use signal processing and machine learning to detect pitch, rhythm, and other musical elements in audio.
    \item The resulting MIDI data can be edited, quantized, or used with virtual instruments.
\end{itemize}

\textbf{MIDI-to-MIDI Generation:}
\begin{itemize}
    \item Tools like Google's Magenta Studio (Continuations), OpenAI's MuseNet, and DeepMind's MusicVAE offer this functionality.
    \item These models can extend MIDI sequences, generate variations, or create accompaniments based on input MIDI data.
    \item Useful for developing musical ideas, creating countermelodies, or generating drum patterns.
\end{itemize}

\textbf{Text-to-MIDI Generation:}
\begin{itemize}
    \item While less common, tools like MuseNet and some iterations of MusicLM can generate MIDI data from text descriptions.
    \item This allows for more precise control over the musical structure compared to direct audio generation.
\end{itemize}

\textbf{Stem Separation:}
\begin{itemize}
    \item Tools like Demucs, Spleeter, and iZotope RX offer this functionality.
    \item These use advanced signal processing and deep learning to isolate individual instruments or vocal tracks. Typically, songs are split into drums, bass, vocals, and other (instruments/other).
    \item Useful for remixing, remastering, or isolating specific elements for further processing.
\end{itemize}

\textbf{Lyric Generation:}
\begin{itemize}
    \item AI language models like ChatGPT, as well as specialized tools like Stacco and Moises AI (used by Strawberry Jam), can assist with this.
    \item These tools can generate lyrics based on themes, styles, or existing lyrical fragments.
    \item Some can even consider rhyme schemes, meter, and genre-specific vocabulary.
\end{itemize}

\textbf{AI Mastering Tools:}
\begin{itemize}
    \item Platforms like Landr, iZotope Ozone, and CloudBounce use machine learning for automated mastering.
    \item These tools analyze the audio and apply appropriate EQ, compression, limiting, and stereo enhancement.
    \item While not a replacement for professional mastering, they can provide a quick, polished sound for demos or small-scale releases like the group’s EPs.
\end{itemize}

\rr{\section{Summary of Major AI Tools Used by Each Team}
\label{Appendix:3}}

\rr{\textbf{Team A (P1, P2, P3, P4)}}

\rr{Team A concentrated primarily on melodic and rhythmic elements. They relied extensively on Suno for melody inspiration across multiple tracks. Google Magenta Studio was another key tool for generating instrumental parts, including bass, drums, bells, and piano arrangements. Additionally, they used Musicfy for melody processing and incorporated Soundraw and Samplab to provide further melody inspiration and transformations.}

\rr{\noindent\textbf{Team B (P5, P6, P9)}}

\rr{Team B emphasized vocal-centric AI applications. They used kits.ai extensively to create realistic female voice models and ensemble vocal harmonies. Complementing their vocal arrangements, they incorporated MusicGen specifically for generating guitar parts and MuseNet for additional audio processing tasks.}

\rr{\noindent\textbf{Team C (P7, P8)}}

\rr{Team C employed the most diverse array of AI tools among the groups, reflecting their exploratory approach. They incorporated Natural Reader AI for spoken vocal elements, Suno for melodic content, and AIVA for composition inspiration. For rhythmic and instrumental aspects, they used Google AI Drum Machine for percussion, Google Tone Transfer to generate flute sounds, and MusicGen to assist specifically in drum generation.}

\end{document}